\documentclass[12pt,twoside]{article}
\usepackage[english]{babel}  
\usepackage[T1]{fontenc} 
\usepackage{enumerate}  
\usepackage{graphics,latexsym,amsfonts}       
\usepackage{amssymb,amsthm,hyperref,mathtools}  
\usepackage{mathrsfs} 
\usepackage{float}
\usepackage{ulem}     

\usepackage[displaymath, mathlines]{lineno}  
\usepackage{microtype} 
\usepackage{mathabx}  
\renewcommand{\emph}{\textsl}

\def\reference#1{\href{#1}{Cliquer ici pour voir une r\'ef\'erence.}} 
\usepackage{siunitx}
\usepackage[toc,page]{appendix}
\usepackage{picture}
\usepackage{sectsty} 

\usepackage{setspace}
 \usepackage{centernot,cancel}
\usepackage{listings}
\usepackage{fancyhdr}
\usepackage{multirow}
\usepackage[section]{placeins}
\usepackage{graphics,latexsym,amsfonts}
\usepackage[affil-it]{authblk}
\usepackage{lipsum}

\usepackage{caption}
\sectionfont{\normalfont\scshape\centering\normalsize}
\subsectionfont{\Large\normalfont\centering\bf}
\providecommand{\U}[1]{\protect \rule{.1in}{.1in}}

{\normalfont\itshape}

\author{
{ \rm Davide Carpentiere}\thanks{Email: davide.carpentiere@phd.unict.it. Corresponding author.}\\
 {\footnotesize \vspace{-4mm}Department of Mathematics and Computer Science, University of Catania, Italy}\\ 
{ \rm Stephen Watson}\thanks{Email: swatson@yorku.ca}\\
{\footnotesize Department of Mathematics and
Statistics, York University, Canada}
}
\title{\bf A new proof for the existence of Nash equilibrium\footnote{The authors wish to thank Alfio Giarlotta, Marco LiCalzi, and M. Ali Khan for many useful comments.}}
\date{}

\def\qed{\hfill $\Box$ \\}

\def\MS{\operatorname{\Sigma}} 
\def\PS{\operatorname{S}} 

\newtheorem{theorem}{Theorem}

\newtheorem{lemma}{Lemma}

\newtheorem{definition}{Definition}

\newcommand{\comment}[1]{}

\begin{document}
\maketitle

\begin{abstract}
\noindent
	We present a new proof for the existence of a Nash equilibrium, which involves no fixed point theorem. 
	The self-contained proof consists of two parts.
	The first part introduces the notions of root function and pre-equilibrium. The second part shows the existence of pre-equilibria and Nash equilibria.

\end{abstract}

\section{Introduction}
Nash equilibrium is regarded as one of the most important notions in Game Theory.
The concept dates back to at least Cournout \cite{Cou1838}.
However, its current formalization is due to Nash, whose original proof \cite{Nash1950}, given in 1950, relies on Kakutani's fixed point theorem \cite{Kak1941}.
One year later, Nash \cite{Nash1951} gave a different proof, which uses Brouwer's fixed point theorem \cite{Brouwer1911}.

The self-contained proof here makes no use of fixed point theorems. 
Our proof can be split in two parts.
The first part introduces two new notions: \textit{root function} and \textit{pre-equilibrium}.
A root function is a map from the set of mixed strategy profiles to the set of pure strategy profiles.
A pre-equilibrium is a subset of mixed strategy profiles that generalizes Nash equilibrium.
In the second part, elaborating an argument used by McLennan and Tourky \cite{McLTou2008}, we show that arbitrarily small pre-equilibria always exist.
By means of compactness, we obtain the existence of a Nash equilibrium.


\section{Proof}
Let $I =\{1, \ldots , n\}$ be the finite set of players, and $\PS_i$ the finite set of strategies of player $i$.
We call $\PS_i$ the set of \textit{pure strategies} of player $i$, and $\PS =\prod \PS_i$ the set of pure strategy \textit{profiles} of the game.
A \textit{mixed strategy} for player $i$ is a probability distribution over $\PS_i$.
The set of all mixed strategies for player $i$ is denoted by $\MS_i$.
Similarly, we define the set of all mixed strategy profiles of the game as $\MS = \prod \MS_i$.
Upon identifying every $\MS_i$ with a $(\vert \MS_i \vert-1)$-simplex, $\MS$ is the product of $n$ simplices.
Therefore we can assume that $\MS$ is contained in some $\mathbb{R}^k$ with the usual distance $d$.

We identify $s \in \PS_i$ with the degenerate probability distribution in $\MS_i$ that assigns probability one to the strategy $s$.
For each player $i$, a \textit{payoff function} is any $g_i \colon \PS \to \mathbb{R}$.
We linearly extend $g_i$ to $f_i \colon \MS \to \mathbb{R}$.
A \textit{Nash equilibrium} is a mixed strategy profile $\sigma \in \MS$ such that for all $i \in \{1,\ldots,n\}$, there is no $\sigma'\in \MS$ with $\sigma_j=\sigma'_j$ for all $j\neq i$, and $f_i(\sigma')>f_i(\sigma)$.

For each mixed strategy $\sigma_i \in \MS_i$ and pure strategy $s \in \PS_i$, we denote by $\alpha_s(\sigma_i)$ the probability of choosing $s$.
We call $\MS$ together with the extended payoff functions $f_i$ a \textit{game}.

\begin{definition} \rm \label{DEF:arrow_notion}
	Suppose $\sigma=(\sigma_i)_{i \in I} \in \MS$ and $i \in I$.
	Define $\sigma(i,s)\in \MS$ by changing strategy $\sigma_i \in \MS_i$ to $s \in \PS_i$.
	Formally, $(\sigma(i,s))_j=\sigma_j$ for all $j \neq i$, and $(\sigma(i,s))_i=s$. 
	Let $A^s_i(\sigma)=\max\{f_i(\sigma(i,s))-f_i(\sigma),0\}$, $A_i(\sigma)=\max\{A_i^s(\sigma) : s \in \PS_i\}$, and $T(\sigma)=\sum_{i \in I}A_i(\sigma)$.
	
	We write $i \Uparrow \sigma$ whenever $A_i(\sigma)> \frac{T(\sigma)}{n+1}$.
\end{definition}

The value $A^s_i(\sigma)$ encodes the possible increase in the payoff of player $i$, in case they change their strategy to the pure strategy $s$.
Furthermore, $A_i(\sigma)$ is the best increase that $i$ can achieve by switching to a pure strategy.
The sum of all possible best increases is $T(\sigma)$.
Note that $T(\sigma)=0$ if and only if $\sigma$ is an equilibrium. 
Whenever $i \Uparrow \sigma$ for some $\sigma$, intuitively, player $i$ gets a good share of $T(\sigma)$.

\begin{definition}\rm \label{DEF:root_function_and_motion}
Define $r: \MS \to \PS$ so that for each $\sigma \in \MS$ and $i \in I$, $r_i(\sigma)=s \in \operatorname{supp}(\sigma_i)$ and $A_i^s(\sigma)=0$.
	Define a function $h \colon \MS \times [0,1] \to \MS$ by setting $h(\sigma,0)=\sigma$, $h(\sigma,1)=r(\sigma)$, and obtaining $h(\sigma,t)$ by linear interpolation for all $t \in (0,1)$, we say that $r$ is a \textit{root function}, and $h$ a \textit{root motion}.
\end{definition}

\begin{lemma}
	For each game there is a root function.
\end{lemma}

\begin{proof}
By contradiction, suppose there is a game, $\sigma \in \MS$, and a player $i$ such that for all $s \in S_i$ the implication $\alpha_s(\sigma_i)>0 \implies f_i(\sigma(i,s))> f_i(\sigma)$ holds.
By minimality, pick a game and a $\sigma \in \MS$ such that $\vert S_i \vert$ is minimal.
If $\vert S_i \vert =1$, then we obtain a contradiction, because  $\sigma(i,s) = \sigma$.
Assume $\vert S_i \vert >1$ and choose $s^*\in S_i$ with $\alpha_{s^{*}}(\sigma_i)\neq 1$ and $\alpha_{s^{*}}(\sigma_i)\neq 0$.

Let $\beta=\alpha_{s^{*}}(\sigma_i)$ and for all $t \in [0,1]$ define $(\sigma_t)_j=\sigma_j$ for all $j \neq i$, $\alpha_s((\sigma_t)_i)=\frac{t}{1-\beta}\alpha_s(\sigma_i)$ for all $s \in S_i\setminus\{s^*\}$, and $\alpha_{s^*}((\sigma_t)_i)=1-t$.
Note that $\sigma_t$ forms a line segment and, by linearity of $f_i$, we obtain $f_i(\sigma_{1-\beta})$ between $f_i(\sigma_0)$ and $f_i(\sigma_1)$.
Since $\sigma_0=\sigma(i,s^*)$, $\sigma_{1-\beta}=\sigma$, and $f_i(\sigma(i,s^*))> f_i(\sigma)$, we obtain $f_i(\sigma_1)\leq f_i(\sigma)$.

To conclude the proof, note that the implication at the start now holds for $\sigma_1$, $i$, and $S_i \setminus\{s^*\}$, which is a contradiction to minimality.
	\qed
\end{proof}

The next lemma collects a few properties satisfied by any root function.

\begin{lemma}\label{LEMMA:root_function_preserves_zeros}
    Let $r$ be a root function.
    We have:
    \begin{itemize}
    	\item [\rm(i)] $\alpha_{r_{i}(\sigma)}(\sigma_i)>0$ and $A_{i}^{r_{i}(\sigma)}(\sigma)=0$;
    	\item [\rm(ii)] $s \notin \operatorname{supp}(\sigma_i) \implies s \notin \operatorname{supp}(r_i(\sigma))$;
    	\item [\rm(iii)] if $\sigma^n \to \sigma \in \MS$, $i \Uparrow \sigma^n$, and $A_i(\sigma)=0$, then $T(\sigma)=0$.
    \end{itemize}
\end{lemma}

\begin{proof}
	Parts (i) and (ii) are straightforward, so we only prove (iii).
    Toward a contradiction, suppose $A_i(\sigma)=0$ but $T(\sigma)>0$.
    By the continuity of $f$, there is an open neighborhood $U$ of $\sigma$ such that $A_i(\sigma')<\frac{T(\sigma')}{n+1}$ for all $\sigma' \in U$.
    It follows that $i \,\cancel\Uparrow \, \sigma'$ for all $\sigma' \in U$, which contradicts $i \Uparrow \sigma^n$ and $\sigma^n \to \sigma$.
    \qed
\end{proof}

Now we define the key new concept of this paper:

\begin{definition}\label{DEF:pre_equilibrium} \rm 
	Let $R \subseteq \MS$.
	We say that $R$ is a \textit{pre-equilibrium} if there is a root function $r$ such that $r(R)=\PS$.
	\end{definition}

We show that pre-equilibria always exist, even arbitrarily small ones. The proof originates in \cite{McLTou2008}.

\begin{lemma}\label{LEMMA:small_pre_equilibrium}
There are arbitrarily small pre-equilibria.
That is, for all $\epsilon >0$ there is a pre-equilibrium $R$ whose diameter is smaller than $\epsilon$.
\end{lemma}

\begin{proof}
   Partition each $\Sigma_j$ into a family of simplices $\mathscr{F}_j$.
   Consider the partition of $\MS$ into the family of simplices $\mathscr{F}=\{F_1 \times \ldots \times F_n : (\forall j \in I)\; F_j \in \mathscr{F}_j\}$.
    Select each $\mathscr{F}_j$ such that the diameter of each $R \in \mathscr{F}$ is smaller than $\epsilon$.
    
	Choose a root function $r$ and let $h: \MS \times [0,1] \to \MS$ be the root motion.
	Identify each $R \in \mathscr{F}$ with its set of vertices and for each $t \in [0,1]$, consider the simplex whose vertices are $h(R,t)$.
	For each $t \in [0,1]$, the volume of each simplex can be expressed as a polynomial in $t$.
	Let $g(t)$ be the sum of the products of these polynomials.
	Note that the volume of $\MS$ is $g(0)$.
	
	By Lemma~\ref{LEMMA:root_function_preserves_zeros}(ii) we obtain that the root motion preserves vertices, edges, and faces of $\MS$.
	Therefore $g(t)$ is constant on some $[0,\delta[$.
	Thus $g$ is a constant polynomial, and $g(1)=g(0)>0$.
	We conclude that there is a simplex $R$ in $\mathscr{F}$ such that $r(R)$ has positive volume.
	Hence $r(R)=\PS$ and $R$ is a pre-equilibrium.
	\qed
\end{proof}

Finally, we have: 

\begin{theorem}\label{THM:theorem}
	There is a Nash equilibrium.
\end{theorem}

\begin{proof}
	Let $\{R_m\}_{m \in \mathbb{N}}$ be a sequence of pre-equilibria with arbitrarily small diameter.
	By compactness, and without loss of generality, $R_m \to \sigma$ for some $\sigma \in \MS$.
	
	Suppose that for infinitely many $m\in \mathbb{N}$ it is true that for all $i \in I$ there is a $\sigma^{i,m} \in R_m$ such that $i \,\cancel\Uparrow \, \sigma^{i,m}\,$ holds.
	Therefore, for infinitely many $m \in \mathbb{N}$, we have $A_i(\sigma^{i,m})\leq \frac{T(\sigma^{i,m})}{n+1}$.
	By continuity $A_i(\sigma)\leq \frac{T(\sigma)}{n+1}$, and summing over $i \in I$ yields $T(\sigma)=0$.
	Hence $\sigma$ is an equilibrium.
	
	Otherwise, for all except a finite number of $m \in \mathbb{N}$, there is an $i \in I$ such that $i \, \Uparrow \, \sigma$ for all $\sigma \in R_m$.
	Without loss of generality fix $i \in I$ so that for all $m \in \mathbb{N}$ and for all $\sigma \in R_m$ we have $i \, \Uparrow \, \sigma$.
	Since $R_m$ is a pre-equilibrium, there is a root function $r^m$ such that $r^m(R_m)=S$.
	Hence for all $s \in S_i$ and for all $m \in \mathbb{N}$ there is $\sigma^{s,m}\in R_m$ such that $r_i^m(\sigma^{s,m})=s$, which implies that $A_i^s(\sigma^{s,m})=0$ by Lemma~\ref{LEMMA:root_function_preserves_zeros}(i).
	For all $s \in \PS_i$, $\sigma^{s,m}\to \sigma$, and so $A_i^s(\sigma)=0$ by continuity.
	We conclude that $A_i(\sigma)=0$ and, by Lemma~\ref{LEMMA:root_function_preserves_zeros}(iii), $T(\sigma)=0$.
	Hence $\sigma$ is an equilibrium.
	\qed
\end{proof}

\end{document}